\documentclass[aps,prl,floats,
  twocolumn,showpacs,superscriptaddress,reprint]{revtex4-1}
\usepackage{graphicx,epsfig}
\usepackage{graphics,dcolumn,bm,epic, eepic,fleqn,float}
\usepackage{amssymb,amsmath,amsfonts,multirow,rotate,color}
\usepackage{soul}
\usepackage{wasysym}

\definecolor{Myorange}{cmyk}{0,0.42,1,0}

\newcommand{\avg}[1]{\langle #1 \rangle}

\newcommand{\lay}[1]{^{[#1]}}
\newcommand{\ud}{\,\mathrm{d}}

\def\be{\begin{equation}}
\def\ee{\end{equation}}

\def\bc{\begin{center}} 
\def\ec{\end{center}}
\def\bea{\begin{eqnarray}}
\def\eea{\end{eqnarray}}

\begin{document}

\title{Growing multiplex networks}

\author{V. Nicosia}
\affiliation{School of Mathematical Sciences, Queen Mary University of
  London, Mile End Road, E1 4NS, London (UK)}
\author{G. Bianconi}
\affiliation{School of Mathematical Sciences, Queen Mary University of
  London, Mile End Road, E1 4NS, London (UK)}
\author{V. Latora}
\affiliation{School of Mathematical Sciences, Queen Mary University of
  London, Mile End Road, E1 4NS, London (UK)}
\affiliation{Dipartimento di Fisica e Astronomia, Universit\`a di Catania and
  INFN, 95123 Catania, Italy}
\author{M. Barthelemy}
\affiliation{Institut de Physique Th\'{e}orique, CEA, CNRS-URA 2306, F-91191, 
Gif-sur-Yvette, France}

\begin{abstract}
  We propose a modelling framework for growing multiplexes where a
  node can belong to different networks. We define new measures for
  multiplexes and we identify a number of relevant ingredients for
  modeling their evolution such as the coupling between the different
  layers and the distribution of node arrival times. The topology of
  the multiplex changes significantly in the different cases under
  consideration, with effects of the arrival time of nodes on the
  degree distribution, average shortest path length and
  interdependence.
\end{abstract}


\maketitle

Many different physical, biological and social systems are structured
as networks, and their properties are now, after a decade of efforts,
well
understood~\cite{Barabasi2002rev,Newman2003rev,Boccaletti2006,Barrat2008}.
However, a complex network is rarely isolated, and some of its nodes
could be part of many graphs, at the same time. Examples include
multimodal transportation networks~\cite{Kurant2006,Zou2010}, climatic
systems~\cite{Donges2011}, economic markets~\cite{Yang2009},
energy-supply networks~\cite{Buldyrev2010} and the human
brain~\cite{Bullmore2009}. In these cases, each network is part of a
larger system in which a set of interdependent networks with different
structure and function coexist, interact and coevolve.
So far network scientists have investigated these systems by looking
at one type of relationship at a time, e.g., by analyzing
collaboration networks and email communications as separate
graphs. However, the structural properties of each of these networks
and their evolution can depend in a non-trivial way on that of other
graphs to which they are interconnected. Consequently, these systems
are better represented as \textit{multiplexes}, i.e. graphs composed
by $M$ different layers in which the same set of $N$ nodes can be
connected to each other by means of links belonging to $M$ different
classes or types. Despite some early attempts in the field social
network analysis~\cite{Wasserman1994}, the characterization of
multiplexes is still in its infancy, mainly due to the lack of
multiplex data. However, some recent works have already proposed
suitable extensions to multi-layer graphs of classic network metrics
and models~\cite{Morris2012,Lee2012,Bianconi2013}. Preliminary results
show that multiplexicity has important consequences for the dynamics
of processes occurring in real systems, including
routing~\cite{Zhuo2011,Morris2012}, diffusion~\cite{Gomez2013},
cooperation~\cite{Gomez-Gardenes2012}, election
models~\cite{Halu2013}, and epidemic
spreading~\cite{Saumell-Mendiola2012}. Nowadays, an increasing number
of new data sets of multiplex systems, e.g. coming from large online
social networks~\cite{Szell2010,Magnani2013}, trading
networks~\cite{Barigozzi2010} and human neuroimaging
techniques~\cite{Nicosia2013}, are rapidly becoming available and
demand for adequate models to understand their structure and
evolution.

In this Letter we propose and study a generic model of multiplex
growth, inspired by classical models based on preferential attachment,
in which the probability for a newly arrived node to establish
connections to existing nodes in each of the layers of a multiplex is
a function of the degree of other nodes at all layers. We define two
new metrics to characterize the structure of multiplexes and we study
the effect of different attachment rules and the impact of delays in
the arrival of nodes at different layers on the structure of the
resulting network. We provide closed forms for both the degree
distributions at each layer and the inter-layer degree-degree
correlations, and we show how different attachment kernels can change
the distributions of distances and  interdependence.

More precisely, a multiplex is a set of $N$ nodes which are connected
to each other by means of edges belonging to $M$ different classes or
types. We represent each class of edges as a separate \textit{layer},
and we assume that a node $i$ of the multiplex consists of $M$
replicas, one for each layer. We denote by $V\lay{\alpha}$ the set of
nodes in layer $\alpha$ and by $E\lay{\alpha}$ the set of all the
edges of a given type $\alpha$. An $M$-layer multiplex is therefore
fully specified by the vector
$\mathcal{A}=[A\lay{1},A\lay{2},\ldots,A\lay{M}]$, whose elements are
the adjacency matrices $A\lay{\alpha}=\{a_{ij}\lay{\alpha}\}$, where
$a_{ij}\lay{\alpha}=1$ if node $i$ and node $j$ are connected by an
edge of type $\alpha$, whereas $a_{ij}\lay{\alpha}=0$ otherwise. We
denote by $k_i\lay{\alpha}=\sum_{j}a_{ij}\lay{\alpha}$ the degree of
node $i$ at layer $\alpha$, i.e. the number of edges of type $\alpha$
of which $i$ is an endpoint, and by $\bm{k}_i$ the $M$-dimensional
vector of the degrees of the replicas of $i$. In general, the degrees
of the replicas of $i$ are distinct, and some replicas can also be
isolated (i.e. $k_i\lay{\alpha}=0$ for some value of $\alpha$). In the
following we consider all the edges at all layers to be undirected and
unweighted. As in the case of classical `singlex' graphs, we can
characterize each layer $\alpha$ of a multiplex by studying the degree
distribution $P(k\lay{\alpha})$, and the joint-degree distribution
$P(k\lay{\alpha},{k'}\lay{\alpha})$. However, we are interested here
in the structural properties of the multiplex as a whole, so we
propose to quantify the correlations between the degrees of replicas
of the same node at two different layers $\alpha$ and $\alpha'$, by
constructing the inter-layer joint-degree distributions
$P(k\lay{\alpha},k\lay{\alpha'})$, or the conditional degree
distributions $P(k\lay{\alpha'} | k\lay{\alpha})$. In particular, we
can look at the projection of the conditional distribution obtained by
considering the average degree $\bar{k}\lay{\alpha'}$ at layer
$\alpha'$ of nodes having degree $k\lay{\alpha}$ at layer $\alpha$:
\begin{equation}
 \bar{k}\lay{\alpha'} (k\lay{\alpha}) = \sum_{k\lay{\alpha'} }
 k\lay{\alpha'} P(k\lay{\alpha'} | k\lay{\alpha})
\end{equation}
By plotting this quantity as a function of $k\lay{\alpha}$ we can
detect the presence and the sign of degree correlations between the
two layers. For a multiplex with no correlations between layers
$\alpha$ and $\alpha'$ we expect $\bar{k}\lay{\alpha'} (k\lay{\alpha})
= \avg{k\lay{\alpha'}}$ and $\bar{k}\lay{\alpha}
(k\lay{\alpha'})=\avg{k\lay{\alpha}}$. If $\bar{k}\lay{\alpha'}
(k\lay{\alpha})$ increases with $k\lay{\alpha}$ we say that the
degrees of the two layers have positive (assortative) correlations,
while if $\bar{k}\lay{\alpha'} (k\lay{\alpha})$ is a decreasing
function of $k\lay{\alpha}$ we say that the degrees on layer $\alpha$
and $\alpha'$ are anticorrelated (or disassortatively correlated). We
notice that a similar concept of inter-network assortativity was
already defined in Ref.~\cite{Parshani2010a} for the case of
interdependent graphs, while the authors of Ref.~\cite{Lee2012}
proposed to measure inter-layer assortativity by means of the
Pearson's linear correlation coefficient of
degrees~\cite{footnote_pearson}.

\begin{figure}[!htbp]
  \begin{center}
    \includegraphics[width=3.1in]{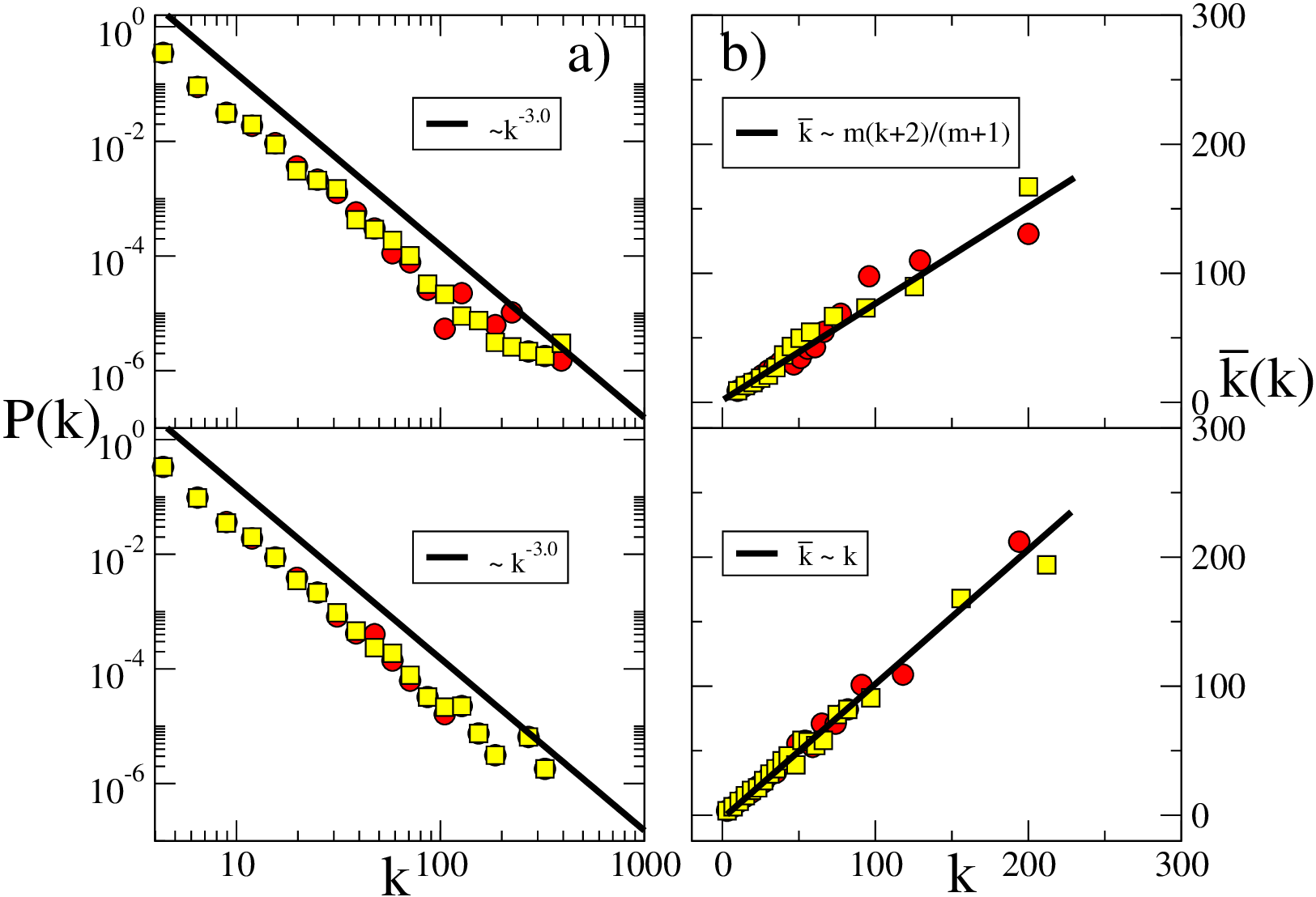}
  \end{center}
  \caption{(color online) \textbf{Synchronous linear attachment.}
    Panel a-b: the degree distribution $P(k)$ (left) and the
    projection $\bar{k}(k)$ of the inter-layer degree correlations
    (right) closely follow the theoretical curves (solid black lines)
    and are relatively insensitive to the coupling matrix.}
  \label{fig:fig1}
\end{figure}

In addition to the assortativity, we can also characterize the
`multiplex reachability' of a node $i$, e.g., by computing the average
distance $L_i$ from $i$ to any other node of the multiplex, and
comparing this average distance with that measured on each layer
separately. The presence of more than one layer in a multiplex
produces an increase in the number of available paths, so that the
distance between two nodes of a multiplex will be, in general, smaller
than or at most equal to that measured on each layer separately. A
better measure to quantify the value added by the multiplexicity to
the reachability of nodes is the
\textit{interdependence}~\cite{Morris2012} which for a node $i$ is
defined by
\begin{equation}
  \lambda_i=\sum_{\stackrel{j\in N}{j\neq i}}
  \frac{\psi_{ij}}{\sigma_{ij}}
\end{equation}
where $\psi_{ij}$ is the number of shortest paths between node $i$ and
node $j$ which use edges lying on more than one layer, while
$\sigma_{ij}$ is the total number of shortest paths between $i$ and
$j$ in the multiplex. The interdependence of a multiplex is computed
as the average node interdependence $\lambda=1/N \sum_{i}\lambda_i$
with $\lambda \in [0,1]$. If $\lambda$ is close to zero, then most of
the shortest paths among nodes lie on just one layer, while if
$\lambda$ is close to $1$ the majority of the shortest paths exploit
more than one layer.

The few models of multiplexes proposed so far are based on the
juxtaposition of random graphs~\cite{Lee2012}. However, networks
usually result from a growing process consisting in the addition of
nodes and edges over time. For this reason, we introduce here a model
of growing multiplex networks. Most of the classical growing models
for single-layer networks start from an initial connected graph with
$m_0$ nodes and assume that new nodes arrive in the graph one by one,
carrying $m$ edge stubs, and connect with other existing nodes
according to a prescribed attachment rule. In that case, each node $i$
has a unique \textit{arrival time} $t_i$, but in multiplexes, instead,
each layer can exhibit a different edge-formation dynamics, and in
general the edges of the $M$ replicas of a new node are not created at
the same time.  For instance, a face-to-face interaction relationship
is usually established before two individuals become friend, while two
locations are usually connected by a road before a direct railway line
between them is constructed. Consequently, we assume that a newly
arrived node has exactly $m$ stubs on each layer of the multiplex (in
Appendix we briefly discuss the case where $m$ is a random variable),
but the replica of a node $i$ on layer $\alpha$ can connect its $m$
stubs at a different time $t_i\lay{\alpha}$. We denote by $\bm{t}_i$
the vector of arrival times of the replicas of node $i$. In order to
make the model analytically tractable, we make two simplifying
assumptions. The first is that there exists a layer $\bar{\alpha}$ so
that $t_i\lay{\bar{\alpha}}\le t_i\lay{\alpha}\>\forall i,\forall
\alpha\neq{\bar{\alpha}}$. This is equivalent to saying that a newly
arrived node must first create its connections on layer $\bar{\alpha}$
before any of its replicas can create connections on any other layer
$\alpha\neq\bar{\alpha}$. We call $\bar{\alpha}$ the \textit{master
  layer} (in Appendix we briefly discuss the case in which this
assumption does not hold, and each node can arrive first on any of the
$M$ layers of the multiplex). The second assumption is that nodes
arrive one by one on the master layer, at equal discrete time
intervals $t=\{1,2,\ldots,\}$. We label the nodes of a growing
multiplex according to the ordering induced by their arrival on the
master layer. Without loss of generality, in the following we assume
that the the master layer is the first one, i.e. $\bar{\alpha}=1$, and
that the arrival times of the replicas of node $i$ have the form
\begin{equation}
  t_{i}\lay{\alpha} = T(t_{i}\lay{1}, \xi\lay{\alpha}(\tau))
  \label{eq:arrival}
\end{equation}
where $T$ is a certain function of $t_{i}\lay{1}$ and of the random
variable $\xi\lay{\alpha}(\tau)$. By appropriately choosing $T$ and
$\xi\lay{\alpha}(\tau)$ we can model different arrival behaviors,
including \textit{a)} \textit{simultaneous arrival}
($T=t_{i}\lay{1}$); \textit{b)} \textit{power-law delayed arrival} ($T
= t_{i}\lay{1} + \xi(\tau)$ and $P(\xi=\tau)=(\beta-1)\tau^{-\beta}$
for $\tau \geq 1$ and $\beta>1$).  Upon
arrival, the newborn node $i$ connects to $m$ existing nodes in the
master layer, according to a certain attachment rule. As in the
preferential attachment models~\cite{Albert1999}, we assume that the
attachment probability depends on the degree of a node. However, in a
multiplex the probability for node $i$ to connect to node $j$ on each
layer $\alpha$ can depend not only on $k_j^{[\alpha]}$ but also on the
degrees of $j$'s replicas on the other layers
\begin{equation}
  \Pi_{i\rightarrow j}\lay{\alpha} =
  \frac{F\lay{\alpha}_j(\bm{k}_j)}{\sum_{l}F\lay{\alpha}_{l}(\bm{k}_{l})}
\end{equation}
For the sake of clarity and without loss of generality, we focus in
the following on 2-layer multiplexes with $\alpha=1,2$.
\begin{figure}[!tbp]
  \begin{center}
    \includegraphics[width=3.1in]{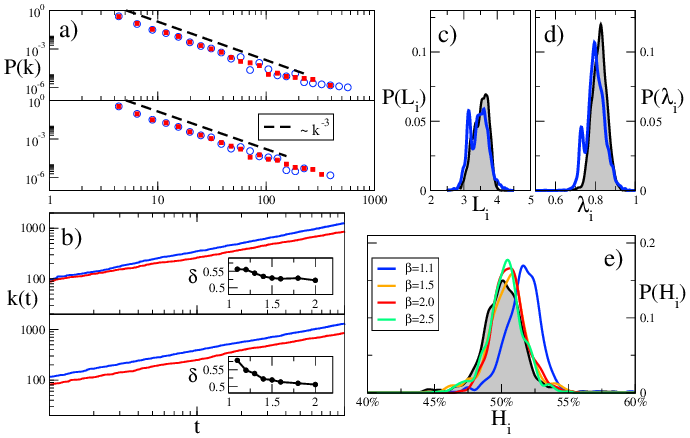}
  \end{center}
  \caption{(color online) \textbf{Delayed linear attachment.}  (a)
    Degree distribution on the first layer. When $\beta$ is close to
    $1$, super-hubs appear. (b) The degree $k(t)$ of the largest hub
    of the first layer as a function of time scales as
    $(t/t_0)^{\delta}$, where $\delta$ approaches $0.5$ when $\beta$
    increases (insets). The value of $\beta$ tunes the shape of the
    distribution of average shortest path lengths (c) and node
    interdependence (d). (e) The percentage of times $H_i$ that the
    maximal-degree node in a shortest path from node $i$ belongs to
    the first layer. When $\beta$ is small, super-hubs in the first
    layer are more abundant in the shortest paths.}
  \label{fig:fig2}
\end{figure}
We begin with the simplest case of \textit{linear attachment} which is
the natural extension of the Barab\'asi-Albert model
\cite{Albert1999}. In this case, we consider that the probability for
a newborn node $i$ to connect to an existing node $j$ on layer
$\alpha$ is proportional to a linear combination of the degrees of $j$
at all layers. The attachment kernels can then be expressed as
\begin{equation}
  \left[\!
  \begin{array}{c}
    F\lay{1}[k,q] \\ 
    F\lay{2}[k,q]
  \end{array}\!\!
  \right]\!\!= C \left[\!\!
  \begin{array}{c}
    k \\
    q
  \end{array}\!\!
  \right]\!\!=\!\! \left[\!\!
  \begin{array}{cc}
    c\lay{1,1} & c\lay{1,2} \\ c\lay{2,1} & c\lay{2,2}
  \end{array}\!\!
  \right]\!\!\!\left[\!\!
  \begin{array}{c}
    k \\
    q
  \end{array}\!\!
  \right]
\end{equation}
where we use here and in the following the notations $k\lay{1}=k$ and
$k\lay{2}=q$~\cite{footnote0}. The coefficients $c\lay{r,s}$ tune the
dependence of the attachment probability at layer $r$ on the degrees
of nodes at layer $s$. In the case of 2-layer multiplexes we can
represent the set of coefficients $C = \{c\lay{r,s}\}$ using the
compact notation $\{c\lay{1,1}, c\lay{1,2}, c\lay{2,1}, c\lay{2,2}\}$.
The dynamics can be easily solved in mean-field (see Appendix for
details) and in some specific cases we can fully characterize the
degree correlations within the two different layers by analytically
solving the master-equation. If we denote by $N_{k,q}(t)$ the number
of nodes having, at time $t$, degree $k$ on the first layer and degree
$q$ on the second layer, and by $\Pi_{k,q}\lay{\alpha}$ the
probability that one of these $N_{k,q}(t)$ nodes acquires one of the
$m$ new links on layer $\alpha$ at time $t+1$, the master equation can
be written as~\cite{footnote1}
\begin{equation}
  N_{k,q}(t+1)=N_{k,q}(t) + \mathcal{G} - \mathcal{L}
  \label{eq:master}
\end{equation}
where
\begin{align*}
  \mathcal{G} &= m\left[\Pi_{k-1,q}^{[1]}N_{k-1,q}(t)+
    \Pi_{k,q-1}^{[2]}N_{k,q-1}(t)\right]+\delta_{k,m}\delta_{q,m}
  \nonumber\\ \mathcal{L} &=
  m\left[\Pi_{k,q}^{[1]}+\Pi_{k,q}^{[2]}\right]N_{k,q}(t)
\end{align*}
represent, respectively, the expected increase ($\mathcal{G}$) and the
expected decrease ($\mathcal{L}$) of $N_{k,q}$ at time $(t+1)$.
Assuming that $N_{k,q} = t P(k,q)$ for large $t$, the solution of
Eq.~(\ref{eq:master}) is obtained by solving the corresponding
recursive expression (see Appendix for details). In the following we
summarize the master-equation solution in some particularly
interesting cases.
First of all let us consider simultaneous arrival of the nodes in the
two layers.  If we set $C=\{1,0,0,1\}$ then the attachment probability
at each layer will depend only on the degree of the nodes in the same
layer. In this case the degree distribution in the first layer
reads~\cite{Dorogovtsev2000,Krapivsky2001}
\begin{equation}
  P(k) = \frac{2m(m+1)}{k(k+1)(k+2)}, \quad k>m
\end{equation}
and the degree distribution in the second layer is identically equal.
This distribution goes as $P(k) \sim k^{-\gamma}$ with $\gamma=3$. If
we solve the master-equation for the multiplex evolution we obtain the
analytical expression for the inter-layer joint degree probability
$P(k,q)$
\begin{equation}
  P(k,q)=\frac{2\Gamma(2+2m)\Gamma(k)\Gamma(q)\Gamma(k+q-2m+1)}
  {\Gamma(m)\Gamma(m)\Gamma(k+q+3)\Gamma(k-m+1)\Gamma(q-m+1)}
\end{equation}
The average degree $\bar{k}(q)$ at layer $1$ of nodes having degree
$q$ at layer $2$ reads:
\begin{equation}
  \bar{k}(q) =\frac{m(q+2)}{1+m}
\end{equation}
Notice that even if the two layers grow independently, the
simultaneous arrival introduces non-trivial inter-layer degree
correlations. In fact, in the mean-field approach, the degree of a
node on each layer increases over time as $k_i\lay{\alpha}(t) =m
\left(t/t\lay{\alpha}_i\right)^{1/2}$ (see Appendix for details), so
that the degrees of the two replicas of a node $i$ depend, for large
$t$, only on their arrival time. If both replicas have the same
arrival time, i.e. $t_i\lay{1}=t_i\lay{2}$ then the degree of the two
replicas will be positively correlated. In Fig.~\ref{fig:fig1} we
report the degree distribution and the values of $\bar{k}(q)$ for two
coupling patterns, which are in good agreement with the theoretical
curves~\cite{footnote2}. It is clear from the figure that in the
synchronous arrival case the shape of the coupling matrix is actually
not very relevant and that the value of the degree distribution
exponent and strong assortativity are robust features of these
multiplexes.

If we consider a power-law delayed arrival time on the second layer,
the results are significantly different. In Fig.~\ref{fig:fig2} we
illustrate how the exponent of the delay distribution $\beta$ affects
the structure of the obtained multiplex. The bulk of the degree
distributions are still power laws $P(k)\sim k^{-\gamma}$ with
$\gamma=3$, but the shape of the far tail depends now on $\beta$: for
small $\beta$, a few nodes are predominant and become super-hubs (as
also shown in Fig.~S-1 in Appendix). The average shortest path and the
interdependence are also significantly affected, as shown in
Fig.~\ref{fig:fig2}c-d. In particular, when $\beta$ is closer to one
the presence of more predominant `old' hubs lowers the average
shortest path and the inter-layer assortativity. Moreover, broader
delays cause a lower participation of hubs of the second layer in
shortest paths, as shown in Fig.~\ref{fig:fig2}e.

\begin{figure}[!tbp]
  \centering
  \includegraphics[width=3in]{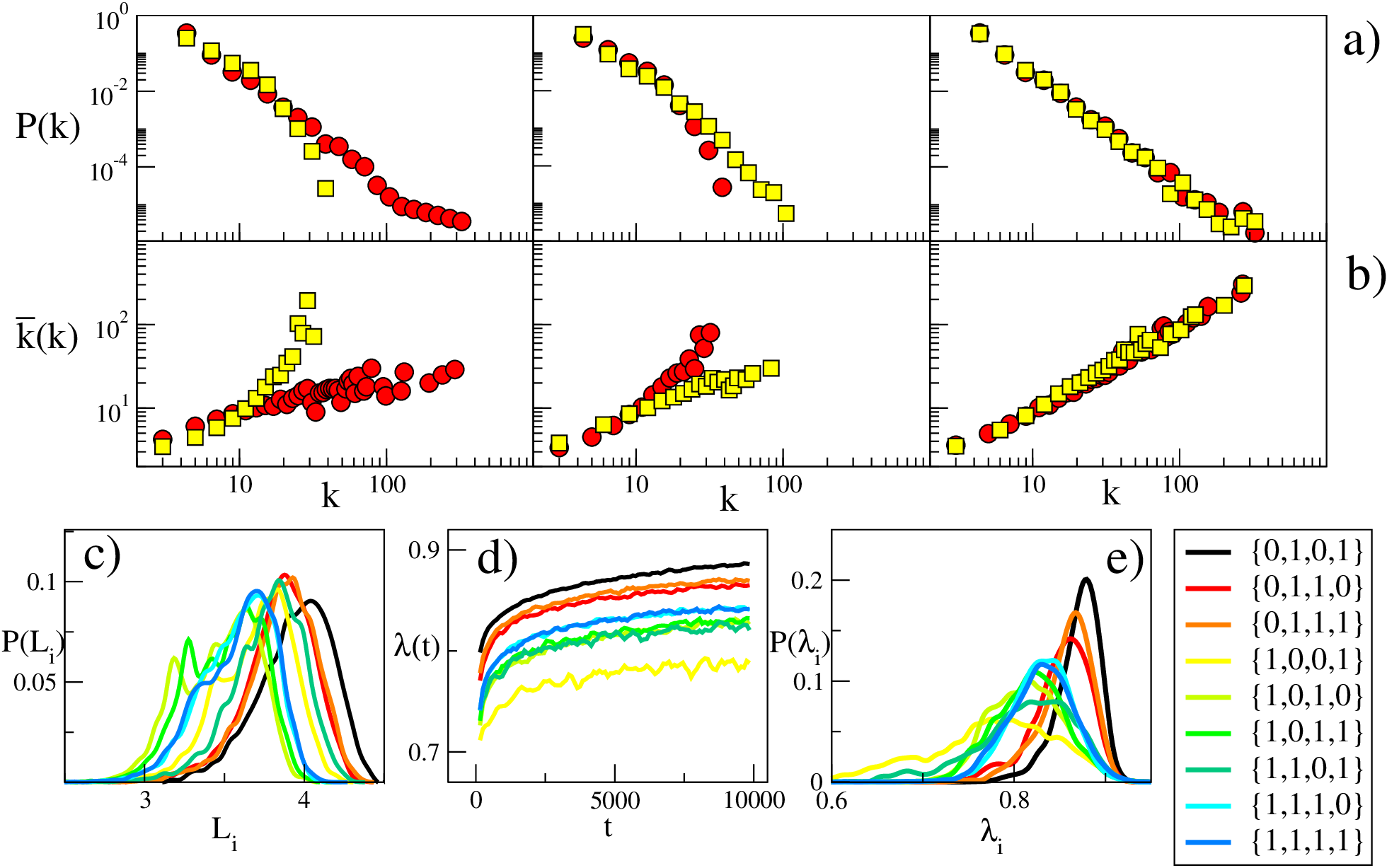}
  \caption{(color online) \textbf{Semi-linear attachment.}  Degree
    distributions (a) and inter-layer degree correlations (b) for
    semi-linear attachment. (c) The distribution of the average
    shortest path length from one node to all the other nodes heavily
    depends on the coupling pattern. Similarly, the interdependence of
    a node $\lambda(t)$ is always a sublinear function of the arrival
    time $t$ but its shape depends on the coupling pattern at work
    (d). In general, older nodes have smaller interdependence. (e) The
    coupling pattern also affects the distribution of node
    interdependence. The smallest average interdependence is observed
    when the two layers are independent (yellow curve).}
  \label{fig:fig3}
\end{figure}

So far we have considered the case of two scale-free growing networks,
but it would be interesting to construct multiplexes in which a
scale-free network is coupled to a network with a peaked degree
distribution. In this respect, we introduce a \textit{semi-linear
  attachment kernel} which allows to grow multiplexes in which the two
layers have different topological structures. The model is defined as
follows
\begin{equation}
  \left[\!
  \begin{array}{c}
    F\lay{1}[k,q] \\ 
    F\lay{2}[{k,q}]
  \end{array}\!\!
  \right] = C \left[\!\!
  \begin{array}{c}
    k \\
    1
  \end{array}\!\!
  \right]
\end{equation}
where $C$ is still a $2\times 2$ matrix of coefficients, as in the
linear model. In this case, the degree of a node on any of the two
layers could depend only on its degree on layer $1$ and does not ever
depend on its degree on layer $2$. If we set $C=\{1,0,0,1\}$ we can
analytically solve the master equation and the degree distributions of
the two layers read:
\begin{equation}
  P\lay{1}(k) \sim k^{-3}, \quad P\lay{2}(q) \sim e^{-q}
\end{equation}
while the inter-layer joint degree distribution is equal to
{\footnotesize
\begin{equation}
  P(k,q) =
  a(k)\sum_{n=0}^{k-m}\binom{k-m}{n}
  \left(\frac{2m}{2+2m+k-n}\right)^{q-m+1} (-1)^{k-m+n}
\end{equation}
}
where $a(k) = \frac{\Gamma(k)}{\Gamma(m+1)\Gamma(k-m+1)}$. The
function $\bar{k}(q)$ is given by
\begin{equation}
  \bar{k}(q) = m\left(\frac{2(m+1)}{1+2m}\right)^{q-m+1}
\end{equation}
Similar relations can be derived for the other coupling patterns.  In
Panel a) and b) of Fig.~\ref{fig:fig3} we report the degree
distribution and the value of $\bar{k}(q)$ for three different
coupling patterns, which are in good agreement with the theoretical
curves. In the semi-linear model the coupling pattern has a dramatic
impact on other structural properties of the multiplex such as the
distribution of the average shortest path length from each node and
the distribution of node interdependence. In particular, the
interdependence is smaller for older nodes, and grows sublinearly with
time. This implies that navigation for old nodes is
easier within a single layer while younger nodes will have to resort
to the different layers to reach a target. In addition, a sublinear
growth implies that the system performance increases very slowly.

\begin{acknowledgments}
  V.N. and V.L. acknowledge support from the Project LASAGNE, Contract
  No.318132 (STREP), funded by the European Commission. M.B. thanks
  Queen Mary University of London for its warm welcome at the start of
  this project, and is supported by the FET-Proactive project PLEXMATH
  (FP7-ICT-2011-8; grant number 317614) funded by the European
  Commission.
\end{acknowledgments}

\clearpage
\renewcommand\theequation{{S-\arabic{equation}}}
\renewcommand\thetable{{S-\Roman{table}}}
\renewcommand\thefigure{{S-\arabic{figure}}}
\setcounter{equation}{0}
\setcounter{figure}{0}
\setcounter{section}{0}

\onecolumngrid
\appendix

\section{Mean-field theory}

In this section, using a mean-field approach, we discuss the time
evolution of the degree of the nodes of a multiplex in the different
layers and we derive long-time expressions for the degree distribution
at each layer and for the inter-layer degree-degree correlations. We
first consider the linear attachment case (A) and then proceed to the
semi-linear case (B). We also provide a concise discussion of the case
in which new nodes bring a random number of new edges (C).

\subsection{Linear attachment kernel on both layers}

According to the growth model discussed in the main text, the
probability that a newly arrived node $i$ in the multiplex creates a
link to node $j$ on layer $\alpha$ can be written as
\begin{equation}
  \Pi_{i\rightarrow j}\lay{\alpha} =
  \frac{F\lay{\alpha}_j(\bm{k}_j)}{\sum_{l}F\lay{\alpha}_{l}(\bm{k}_{l})}
\end{equation}
where $F\lay{\alpha}_j(\bm{k}_j)$ is a certain function of the degrees
of the replicas of node $j$.  If $F\lay{\alpha}_j(\bm{k}_j)$ is a
linear function of $\bm{k}_j$ $\forall \alpha$, and all the replicas
of the new node arrive at the same time, the temporal evolution of the
degree of a node on each layer is governed by the equations
\begin{equation}
  \left[\!
  \begin{array}{c}
   \frac{d{k\lay{1}}}{dt} \\ 
    \frac{d{k\lay{2}}}{dt}
  \end{array}\!\!
  \right]\!\!= \frac{1}{2t}C \left[\!\!
  \begin{array}{c}
    k^{[1]} \\
    k^{[2]} 
  \end{array}\!\!
  \right]\!\!=\!\! \frac{1}{2t}\left[\!\!
  \begin{array}{cc}
    c\lay{1,1} & c\lay{1,2} \\ c\lay{2,1} & c\lay{2,2}
  \end{array}\!\!
  \right]\!\!\!\left[\!\!
  \begin{array}{c}
    k^{[1]} \\
    k^{[2]} 
  \end{array}\!\!
  \right]
  \label{eq:eq1}
\end{equation}
with the constraints $c\lay{1,1}+c\lay{1,2}=1$ and
$c\lay{2,1}+c\lay{2,2}=1$.  Since the matrix elements are real and
non-zeros, the maximal eigenvalue is real. Moreover since we have just
two layers then both eigenvalues $\lambda_1$ and $\lambda_2$ are
real. We notice that if we impose that each row of matrix $C$ must sum
to $1$, and that the coefficients $c\lay{r,s}$ are non-negative (to
ensure that $\Pi\lay{\alpha}_{i\rightarrow j}$ is a probability
distribution $\forall \alpha$), then we can write
\begin{equation*}
  C=\left[\begin{array}{cc} 
      a & 1-a \\ 
      b & 1-b
  \end{array}\!\!\right], \quad   0\le a \le 1, 0\le b \le 1
\end{equation*}
It is easy to verify that if $b-a+1\neq 0$ the matrix $C$ has
eigenvalues $\lambda_1=1$ and $\lambda_2=(a-b)\neq 1$, with
eigenvectors $\bm{u^1} = [1,1]$ and $\bm{u^2} = [1, -b/(1-a)]$ (the
degenerate case $b-a+1=0$ is considered below). Since the eigenvalues
are distinct, then $C$ is diagonalizable, i.e. it is similar to the
diagonal matrix
\begin{equation*}
  \Lambda = 
  \left[\begin{array}{cc} 
      1 & 0 \\ 
      0 & a-b
  \end{array}\!\!\right], \quad   0\le a \le 1,\> 0\le b \le 1
\end{equation*}
whose non-zero elements are the eigenvalues of $C$. The system in
Eq.~(\ref{eq:eq1}) can be also written in the form
\begin{equation}
  \dot{\bm{k}}(t) = A(t)\bm{k}(t) = \alpha(t)C\bm{k}(t)
\end{equation}
where $\alpha(t)=\frac{1}{2t}$ is a scalar function, and $C$ is a
constant matrix. This is a homogeneous time-varying linear dynamical
system, whose temporal evolution is fully determined by the initial
state $\bm{k}_s=m{\bf u^1}$ and by the state transition matrix
$\Phi(t,s)$, where $s$ is the time at which a node is added to the
graph
\begin{equation}
  \bm{k}(t) = \Phi(t,s)\bm{k}_s= m\Phi(t,s){\bf u^1}.
\end{equation}
Since $C$ is diagonalizable, then the transition matrix can be written
as
\begin{equation}
  \Phi(t,s) = e^{\int_s^tA(\tau)\ud \tau}=e^{C\int_s^{t}\frac{\ud \tau}{2\tau}}=e^{C\sigma}.
\end{equation}
To compute the transition matrix $\Phi(t,s)$ we first compute the
exponential matrix $e^{C\sigma}$ and then we substitute $\sigma$ with
$\int_{s}^{t}\frac{\ud \tau}{2\tau}$. Since the eigenvalues of $C$ are
distinct, the corresponding eigenvectors form a base of
$\mathbb{R}^2$, so we have
\begin{equation}
  e^{C\sigma} = V e^{\Lambda}V^{-1}
\end{equation}
where $V$ is the matrix whose columns are the eigenvectors of $C$ and
$\Lambda$ is the diagonal matrix of the eigenvalues of $C$. After some
simple algebra we obtain
\begin{equation}
  e^{C\sigma} = \frac{1}{1-(a-b)}
  \begin{bmatrix}
    be^{\sigma} + (1-a) e^{(a-b)\sigma} & \quad(1-a)e^{\sigma} -
    (1-a)e^{(a-b)\sigma}\\ be^{\sigma}+\frac{b(1-a)}{a-b}e^{(a-b)\sigma}
    & \quad(1-a)e^{\sigma} - \frac{b(1-a)}{(a-b)}e^{(a-b)\sigma}
  \end{bmatrix}.
\end{equation}
By substituting $\sigma$ with $\int_{s}^{t}\frac{\ud
  \tau}{2\tau}=\frac{1}{2}(\log{t}-\log{s})$, we get:
\begin{equation}
  \Phi(t,s) =  \frac{1}{1-(a-b)}\begin{bmatrix} b(\frac{t}{s})^{\frac{1}{2}} +
    (1-a)(\frac{t}{s})^{\frac{a-b}{2}} & \quad
    (1-a)(\frac{t}{s})^{\frac{1}{2}}-
    (1-a)(\frac{t}{s})^{\frac{a-b}{2}}\\ b (\frac{t}{s})^{\frac{1}{2}}
    + \frac{b(1-a)}{a-b}(\frac{t}{s})^{\frac{a-b}{2}} & \quad
    (1-a)(\frac{t}{s})^{\frac{1}{2}} -
    \frac{b(1-a)}{a-b}(\frac{t}{s})^{\frac{a-b}{2}}
    \end{bmatrix}
\end{equation}
and the solution of the system reads
\begin{equation}
  \bm{k}_s(t) = m\Phi(t,s){\bf u^1} =
    m\left(\frac{t}{s}\right)^{\frac{1}{2}}{\bf u^1}.
\end{equation}

In this case we have

\begin{align}
  \avg{k\lay{1}|k{\lay{2}}}\propto k\lay{2},\\
    \avg{k\lay{2}|k{\lay{1}}}\propto k\lay{1}.
\end{align}

Let us now consider the degenerate case $b-a+1=0$. Since we imposed that
each row of the matrix $C$ has to sum to $1$, then $b-a+1=0$ only if
$a=1$ and $b=0$. In this case the two layers evolve independently, the
matrix $C$ is diagonal and the time evolution of the degree on each
layer reads
\begin{equation}
  k_s\lay{\alpha}(t) = m\left(\frac{t}{s}\right)^{\frac{1}{2}},
\end{equation}
with $\alpha=1,2$.
This means that in the case of linear attachment kernel on both layers
without delay the degree distribution of each layer is a power-law
$P(k)\sim k^{-\gamma}$ with exponent $\gamma=3$ and we have
\begin{align}
  \avg{k\lay{1}|k{\lay{2}}}\propto k\lay{2},\\
    \avg{k\lay{2}|k{\lay{1}}}\propto k\lay{1}.
\end{align}

\subsection{Semi-linear attachment kernel}
For the semi-linear attachment kernel we have
\begin{equation}
    \left[\!
  \begin{array}{c}
   \frac{d{k\lay{1}}}{dt} \\ 
    \frac{d{k\lay{2}}}{dt}
  \end{array}\!\!
  \right]\!\!= \frac{1}{t}
    \begin{bmatrix}
      \frac{am}{2am+1-a} & 0 \\
      \frac{bm}{2bm+1-b} & 0
    \end{bmatrix}
    \left[\!\!
  \begin{array}{c}
    k\lay{1} \\
    k\lay{2}
  \end{array}\!\!
  \right] + \frac{1}{t}
    \begin{bmatrix}
      \frac{m(1-a)}{2am+1-a} & 0 \\
      0 & \frac{m(1-b)}{2bm+1-b}
    \end{bmatrix}
    \begin{bmatrix}
      1 \\
      1
    \end{bmatrix}
  \label{eq:eq2}
\end{equation}
which is in the form 
\begin{equation}
  \dot{\bm{k}}(t) = A(t)\bm{k}(t) + B(t) \bm{w}(t)
  \label{eq:complete_system}
\end{equation}
where
\begin{equation*}
  A(t)= 
  \begin{bmatrix}
    \frac{am}{t(2am+1-a)} & 0 \\
    \frac{bm}{t(2am+1-a)} & 0
  \end{bmatrix}, \quad 
  B(t)=
  \begin{bmatrix}
    \frac{m(1-a)}{t(2am+1-a)} & 0 \\
    0 & \frac{m(1-b)}{t(2bm+1-b)}
  \end{bmatrix},
  \quad
  \bm{w}(t) = 
  \begin{bmatrix}
    1 \\
    1
  \end{bmatrix}.
\end{equation*}
The system in Eq.~(\ref{eq:complete_system}) is a non-homogeneous
time-varying linear dynamical system where $\bm{w}(t)$ represents an
external forcing function. If we call $\Phi(t,s)$ the state transition
matrix of the corresponding homogeneous system $\dot{\bm{k}}(t) = A(t)
\bm{k}(t)$, it is possible to show that the unique solution of
Eq.~(\ref{eq:complete_system}) is given by
\begin{equation}
  \bm{k}_s(t) = \Phi(t,s)\bm{k}_s +
  \int_{s}^{t}\!\!\!\ud\sigma\,\Phi(t,\sigma)B(\sigma)w(\sigma).
  \label{eq:gen_solution}
\end{equation}
The form of the transition matrix $\Phi(t,s)$ associated to the
homogeneous system depends on the value of $a$. When $a\neq 0$ then
$\Phi(t,s)$ reads
\begin{equation}
  \Phi(t,s)= \begin{bmatrix}
    \left(\frac{t}{s}\right)^{\beta} &\quad\quad
    0\\ \frac{(2abm+(1-a )b)}{(2
      abm-ab+a)}\left(\frac{t}{s}\right)^{\beta}
    +\frac{((a-1)b-2abm)}{(2 abm-ab+a)} & \quad\quad 1
    \end{bmatrix}, \quad\quad\quad \text{where}\quad\beta=\frac{am}{2am-a+1}.
  \label{eq:phi_ts}
\end{equation}
By plugging Eq.~(\ref{eq:phi_ts}) into Eq.~(\ref{eq:gen_solution}) one
obtains the mean-field temporal evolution of $k_s\lay{1}$ and
$k_s\lay{2}$: 
\begin{align}
  k_s\lay{1}(t) &= \frac{am-a+1}{a}\left(\frac{t}{s}\right)^{\beta} +
  \frac{a-1}{a}\nonumber,\\ k_s\lay{2}(t) &= \delta
  \left(\frac{t}{s}\right)^{\beta} + \eta\left(\log(t)-\log(s)\right)
    +\epsilon,
    \label{eq:semi_time}
\end{align}
where
\begin{align*}
  \delta &= \frac{2a^2bm^2+(3a-3a^2)bm+(a+1)^2b}{2a^2bm-a^2b+a^2},\\
  \eta &= \frac{(a^2-ab)m}{2a^2bm-a^2b+a^2},\\
  \epsilon &= \frac{((2a^2-3a)b + a^2)m - (a-1)^2b}{2a^2bm-a^2b+a^2}.
\end{align*}
In general, if $b\neq 0$ then for $t\rightarrow\infty$ we have
$\left(\frac{t}{s}\right)^{\beta}\gg \left(\log(t) -
\log(s)\right)$. Consequently, Eqs.~(\ref{eq:semi_time}) can be
written as
\begin{align}
  k_s\lay{1}(t) & \simeq \left(m + \frac{1-a}{a}\right)
  \left(\frac{t}{s}\right)^{\beta} \label{eq:semi1},\\ k_s\lay{2}(t) & \simeq
  \delta\left(\frac{t}{s}\right)^{\beta}
  \label{eq:semi_b_neq_0}
\end{align}
so that the degree distribution on both layers reads
\begin{equation}
  P(k\lay{\ell}) \sim k^{-(\frac{1}{\beta}+1)} = k^{-(3
    +\frac{1-a}{am})} 
  \label{eq:pk_semi_b_neq_0}
\end{equation}
and we have
\begin{align}
  \avg{k\lay{1}|k\lay{2}} & \propto k\lay{2}\nonumber, \\
  \avg{k\lay{2}|k\lay{1}} & \propto k\lay{1}.
\end{align}
Instead, if $b=0$ the solution for the degree of nodes  on the second
layer reads
\begin{equation}
  k_s\lay{2}(t) = \eta (\log(t) - \log(s)) + \epsilon,
  \label{eq:pk2_semi_b_0}
\end{equation}
while $k_s\lay{1}(t)$ is expressed by Eq.~(\ref{eq:semi1}).
In this case, the degree distribution on the first layer is the same
as in Eq.~(\ref{eq:pk_semi_b_neq_0}), while for the second layer we
have
\begin{equation}
  P(k\lay{2}) \sim e^{-\frac{k}{\eta}}
\end{equation}
and in the limit of large $k\lay{1}(t),k\lay{2}(t)$ we obtain
\begin{align}
  \avg{k\lay{1}|k\lay{2}} & \propto e^{\frac{\beta k\lay{2}}{\eta}}\nonumber, \\
  \avg{k\lay{2}|k\lay{1}} & \propto \log(k\lay{1}).
\end{align}
Eqs.~(\ref{eq:phi_ts}---\ref{eq:pk2_semi_b_0}) are valid when $a\neq
0$. When $a=0$ the state transition matrix reads
\begin{equation}
  \Phi(t,s) = \begin{bmatrix}
    1 & \quad\quad 0\\
    \frac{bm\left(\log t-\log s\right)}{2bm-b+1} & \quad\quad 1
  \end{bmatrix}
\end{equation}
and the generic solutions for $k_s\lay{1}(t)$ and $k_s\lay{2}(t)$ are
\begin{align}
  k_s\lay{1}(t) & = m \left(\log(t) - \log(s)\right) + m
  \nonumber,\\ k_s\lay{2}(t) &=
  \frac{bm^2\left(\log(t)-\log(s)\right)^2 + \left((2-2b)m +
    2bm^2\right)\left(\log(t)-\log(s)\right) + 4bm^2 + (2-2b)m}{4bm
    -2b +2}.
  \label{eq:semi_time_a_0}
\end{align}
In this case, the degree distribution on the first layer is
exponential $P(k\lay{1})\sim e^{-\frac{k}{m}}$. On the second layer, the
functional form of the degree distribution depends on the value of
$b$. It is easy to verify that when $b=0$ then $P(k\lay{2})\sim
e^{-\frac{k}{m}}$, and we have in the limit of large $k\lay{1}(t),k\lay{2}(t)$
\begin{align}
  \avg{k\lay{1}|k\lay{2}} & \propto k\lay{2}\nonumber, \\
  \avg{k\lay{2}|k\lay{1}} & \propto k\lay{1}.
\end{align}
Conversely, when $b>0$ the degree distribution on the second layer is 
\begin{equation}
  P(k\lay{2})\sim \frac{e^{-\sqrt{\mu k + \nu}}}{\sqrt{\mu k + \nu}}
\end{equation}
where
\begin{align*}
  \mu &= m\left[4b^2m + \left(2b-2b^2\right)\right], \\
  \nu &= m^2\left[b^2m^2 + \left(2b-6b^2\right)m + (3b^2-4b+1)\right].
\end{align*}
In this case, in the limit of large $k\lay{1}(t),k\lay{2}(t)$, we have
\begin{align}
  \avg{k\lay{1}|k\lay{2}} & \propto \sqrt{k\lay{2}}\nonumber, \\
  \avg{k\lay{2}|k\lay{1}} & \propto \left(k\lay{1}\right)^{2}.
\end{align}

\subsection{Fluctuations in the number of edges}

In principle, the mean-field approach could be also applied to the
case in which the number of edges brought on layer $\alpha$ by each
new-born node is not fixed but is a random variable $\xi\lay{\alpha}$
drawn from a given distribution $P(\xi\lay{\alpha})$. In this case we
should solve the system of stochastic differential equations:
\begin{equation}
  \begin{bmatrix}
   \frac{d{k\lay{1}}}{dt} \\ 
    \frac{d{k\lay{2}}}{dt}
    \end{bmatrix}\!\!= 
  \frac{1}{2\left(\kappa\lay{1}(t) + \kappa\lay{2}(t)\right)}
  \begin{bmatrix}
    \xi\lay{1}(t) & 0 \\
    0 & \xi\lay{2}(t) 
  \end{bmatrix}
  \begin{bmatrix}
    a & 1-a\\
    b & 1-b
  \end{bmatrix}
  \begin{bmatrix}
    k\lay{1}\\
    k\lay{2}
  \end{bmatrix}
\end{equation}
where:
\begin{equation*}
  \kappa\lay{\alpha}(t) = \int_{1}^{t}\!\!\!\ud\tau\,\, \xi\lay{\alpha}(\tau)
\end{equation*}
The random variable $\xi\lay{\alpha}$ is a positive integer with
average $\langle\xi\lay{\alpha}\rangle$ and the dominant term at large
times of $\kappa\lay{\alpha}$ is then given by
\begin{equation*}
  \kappa\lay{\alpha}(t) \simeq t\avg{\xi\lay{\alpha}}+o(t)
\end{equation*}
This implies in particular that at large times, the effect of
randomness in the number of edges is negligible, and the behavior of
the system is governed by the average number of edges
$\avg{\xi\lay{\alpha}}$ added in each layer.

\section{Master Equation approach for the model without delay}
We provide here the derivation of exact expressions of $P(k)$ and
$P(k,q)$ starting from the master equation of the system.
We denote by $N_{k,q}(t)$ the average number of nodes that at time $t$
have degree $k$ in layer 1 and degree $q$ in layer 2.  We start from a
small connected network and at each time we add a node which brings,
at same time, $m$ new edges in layer 1 and $m$ new edges in layer
2. We assume that, when we add the new node $i$ to the network, the
expected number of new links in layer 1 attached to a node $j$ of
degree $k$ in layer 1 and degree $q$ in layer 2 is given by
$m\Pi^{[1]}_{i\to j}=\frac{A_{k,q}}{t}$. Similarly, the expected
number of new links in layer 2 attached to a node $j$ of degree $k$ in
layer 1 and degree $q$ in layer 2 is given by $m\Pi^{[2]}_{i\to
  j}=\frac{B_{k,q}}{t}$ .  In addition to that, we work in the
hypothesis that in the large $t$ limit, $t\gg 1$, we have
$A_{k,q}/t\ll1$ and $B_{k,q}/t\ll1$ so that we can neglect the
probability that a node acquires at the same time a link in both
layers.  In this hypothesis the master equation for evolving
multiplex network is given by \bea
N_{k,q}(t+1)=N_{k,q}(t)+\frac{A_{k-1,q}}{t}N_{k-1,q}(t)+\frac{B_{k,q-1}}{t}N_{k,q-1}(t)-\left[\frac{A_{k,q}}{t}+\frac{B_{k,q}}{t}\right]N_{k,q}(t)+\delta_{k,m}\delta_{q,m}
\eea for $k\geq m$ and $q\geq m$, as long as
$N_{m-1,q}(t)=N_{k,m-1}(t)=0$.  Assuming that $N_{k,q}=tP(k,q)$ is
valid in the large time limit $t\gg1 $, we can solve for the combined
degree distribution $P(k,q)$ indicating the probability that a node
has at the same time degree $k$ in layer 1 and degree $q$ in layer
2. We get the master equations 
\bea
P(m,q)&=&\left(\prod_{j=m+1}^{q}\frac{B_{m,j-1}}{1+A_{m,j}+B_{m,j}}\right)P(m,m)\nonumber,
\\ P(k,q)&=&\sum_{r=m}^{q}\left(\prod_{j=r+1}^{q}
\frac{B_{k,j-1}}{1+A_{k,j}+B_{k,j}}\right)\frac{A_{k-1,r}}{1+A_{k,r}+B_{k,r}}P(k-1,r)
\label{sol}
\eea

\subsection{Solution of the master equation in three simple cases}

\textit{(i) Linear attachment kernel -- } Let us first consider a linear preferential
attachment kernel, in which $c^{[1,1]}=c^{[2,2]}=1$ and
$c^{[1,2]}=c^{[2,1]}=0$. In this case we have
\begin{align}
A_{k,q}=\frac{k}{2},\nonumber\\ B_{k,q}=\frac{q}{2}.
\end{align}
The recursive Eqs.~(\ref{sol}) read
\begin{align}
P(m,q)&=\frac{\Gamma(q)\Gamma(3+2m)}{\Gamma(m)\Gamma(3+q+m)}P(m,m)\nonumber,&\\ 
P(k,q)&=\sum_{r=m}^{q}
\left(
\frac{\Gamma(q)\Gamma(3+r+k)}{\Gamma(r)\Gamma(3+q+k)}\right)\frac{k-1}{2+k+q}P(k-1,r)&
\end{align} 
where $P(m,m)$ is fixed by the normalization condition
$\sum_{k=m}^{\infty}\sum_{q=m}^{\infty}P(k,q)=1$.  Using the relation
\begin{align}
\sum_{r=m}^{q}\frac{\Gamma(k+r-2m)}{\Gamma(r-m+1)\Gamma(k-m)}=\frac{\Gamma(k+q-2m+1)}{\Gamma(k-m+1)\Gamma(q-m+1)}
\label{relation}
\end{align}
it can be proved recursively that $P(k,q)$ takes the following
expression 
\begin{align}
P(k,q)=\frac{2\Gamma(2+2m)}{\Gamma(m)\Gamma(m)}\frac{\Gamma(k+q-2m+1)}{\Gamma(k+q+3)}\frac{\Gamma(q)}{\Gamma(q-m+1)}\frac{\Gamma(k)}{\Gamma(k-m+1)}\label{eq:P_kq_diagonal}
\end{align}
Summing over the degree in layer 2 we can find the degree distribution
$P(k)$ in layer 1, i.e. $P(k)=\sum_{q=m}^{\infty}P(k,q)$ obtaining the
known result for a single layer, 
\begin{align} 
P(k)=\frac{2m(1+m)}{k(k+1)(k+2)}
\end{align} 
The function $\avg{k(q)}$ is given by 
\begin{align}
\avg{k(q)}=\frac{\sum_{k=m}^{\infty}kP(k,q)}{\sum_{k=m}^{\infty}P(k,q)}=\frac{m}{1+m}(q+2)
\end{align}
Similar expressions are obtained for $P(q)$ and $\avg{q(k)}$, by
summing Eq.~(\ref{eq:P_kq_diagonal}) over $k$.

\bigskip
\noindent
\textit{(ii) Uniform attachment kernel -- } Let us now consider a
uniform attachment kernel, in which every target node $j$ is chosen
with probability $\Pi^{[1]}_{i \to j}=\frac{1}{t}$ in layer 1 and with
probability $\Pi^{[2]}_{i\to j}=\frac{1}{t}$ in layer 2, so that
\begin{align} A_{k,q}=m\nonumber\\ B_{k,q}=m.  \end{align} 
In this case the recursive Eqs.~(\ref{sol}) read 
\begin{align}
&P(m,q)=\left(\frac{m}{1+2m}\right)^{q-m}P(m,m)\nonumber,&
\\ &P(k,q)=\sum_{r=m}^{q}\left(\frac{m}{1+2m}\right)^{q-r}\frac{m}{1+2m}P(k-1,r)&
\label{sol2}
\end{align}
where $P(m,m)$ is again fixed by the normalization condition
$\sum_{k=m}^{\infty}\sum_{q=m}^{\infty}P(k,q)=1$.  Using again the
relation provided in Eq.~(\ref{relation}) it is easy to prove
recursively that $P(k,q)$ is given in this case by 
\begin{align}
P(k,q)=\frac{1}{m}\left(\frac{m}{1+2m}\right)^{k+q-2m+1}\frac{\Gamma(k+q-2m+1)}{\Gamma(k-m+1)\Gamma(q-m+1)}
\end{align}
Moreover the degree distribution $P(k)=\sum_{q=m}^{\infty}P(k,q)$
and $P(q)=\sum_{k=m}^{\infty}P(k,q)$ of a single network are given by
\begin{align}
P(k)=\frac{1}{1+m}\left(\frac{m}{1+m}\right)^{k-m}\text{ and }\quad
P(q)=\frac{1}{1+m}\left(\frac{m}{1+m}\right)^{q-m}
\end{align}
while the function $\avg{k(q)}=\sum_{k=m}^{\infty}kP(k,q)/P(q)$ is
given by 
\begin{align} \avg{k(q)}=\frac{(q+2)m}{1+m} 
\end{align}

\bigskip
\noindent
\textit{(iii) Semi-linear attachment kernel -- } Finally we analyze
the case of semi-linear attachment with $c^{[1,1]}=c^{[2,2]}=1$ and
$c^{[1,2]}=c^{[2,1]}=0$. We have:
\begin{align} A_{k,q}=\frac{k}{2}\nonumber \\ B_{k,q}=m.  
\end{align} 
In this case the Eqs.~(\ref{sol}) read as 
\begin{align}
&P(m,q)=\left(\frac{2m}{2+3m}\right)^{q-m}P(m,m)\nonumber,&
\\ &P(k,q)=\sum_{r=m}^{q}\left(\frac{2m}{2+k+2m}\right)^{q-r}\frac{k-1}{2+k+2m}P(k-1,r)&
\end{align}
where $P(m,m)$ is fixed by the normalization condition
$\sum_{q=m}^{\infty}\sum_{k=m}^{\infty}P(k,q)=1$.  It can be shown
recursively that these equations have the following solution, 
\begin{align}
P(k,q)=\frac{1}{\Gamma(m+1)}\frac{\Gamma(k)}{\Gamma{(k-m+1)}}\sum_{n=0}^{k-m}\left(\begin{array}{c}k-m\nonumber\\ n\end{array}\right)\left(\frac{2m}{2+2m+k-n}\right)^{q-m+1}(-1)^{k-m-n},
\end{align}
with the associated degree distributions
$P(k)=\sum_{q=m}^{\infty}P(k,q)$ and $P(q)=\sum_{k=m}^{\infty}P(k,q)$
given by 
\bea P(k)&=&\frac{2m(1+m)}{k(k+1)(k+2)}\nonumber,
\\ P(q)&=&\frac{1}{\Gamma{(1+m)}}\sum_{k=m}^{\infty}\sum_{n=0}^{k-m}\left(\begin{array}{c}k-m\nonumber\\ n\end{array}\right)\left(\frac{2m}{2+2m+k-n}\right)^{q-m+1}(-1)^{k-m+n}=\left(\frac{m}{1+m}\right)^{q-m}\frac{1}{1+m}
\eea 
Finally the function $\avg{k(q)}=\sum_{k=m}^{\infty}kP(k,q)/P(q)$ is
given by 
\begin{align}
\avg{k(q)}&=\left(\frac{2(m+1)}{1+2m}\right)^{q-m+1}m\nonumber\\ 
\end{align}

\section{Role of $\beta$ in the delayed arrival}

\begin{figure}[!ht]
  \begin{center}
    \includegraphics[width=5in]{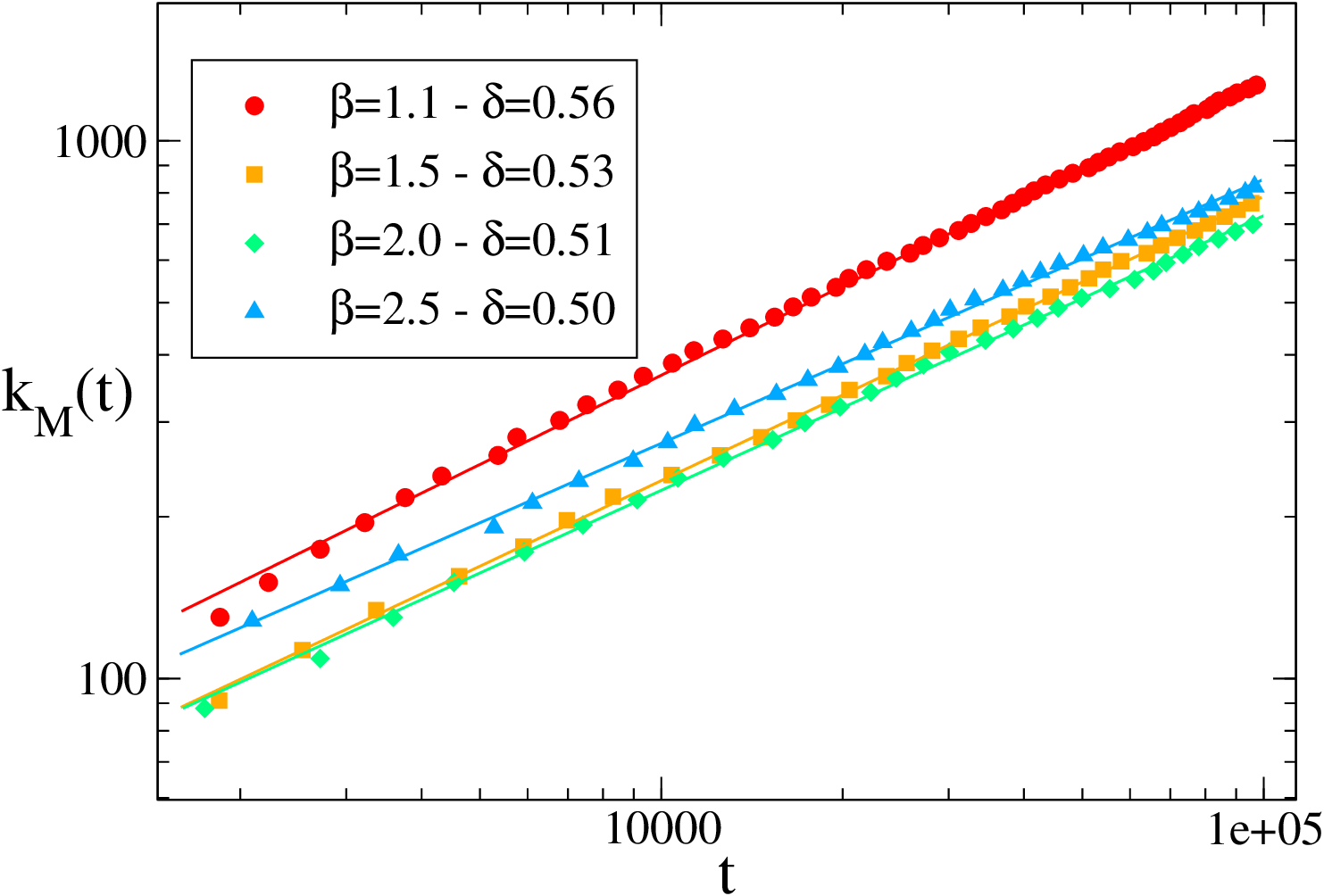}
  \end{center}
  \caption{\textbf{Effect of $\beta$ on the maximum degree of a
      layer.} When the exponent $\beta$ of the delay distribution is
    close to $1$, the maximum degree of a layer scales as
    $(t/s)^{\delta}$, with $\delta>1/2$. Consequently, the rightmost
    side of the degree distribution is broader when $\beta$ is close
    to $1$ and converges to $P(k) \sim k^{-3}$ as $\beta$ increases.}
  \label{fig:suppl_fig_1}
\end{figure}

In Fig.~\ref{fig:suppl_fig_1} we show the time evolution of the
maximum degree $k_{M}(t)$ on the first layer, for different values of
$\beta$. Notice that $k_M(t) \sim (t/s)^{\delta}$. The effect of the
exponent $\beta$ tuning the width of the delay distribution is
evident: the larger the value of $\beta$, the closer $\delta$ is to
$0.5$, the value observed in the case of synchronous
arrival. Consequently, the rightmost part of the degree distribution
is broader when $\beta$ is close to $1$ and becomes more similar to
$P(k) \sim k^{-3}$ when $\beta$ increases.

\section{Finite size effects}

\begin{figure}[!ht]
  \begin{center}
    \includegraphics[width=5in]{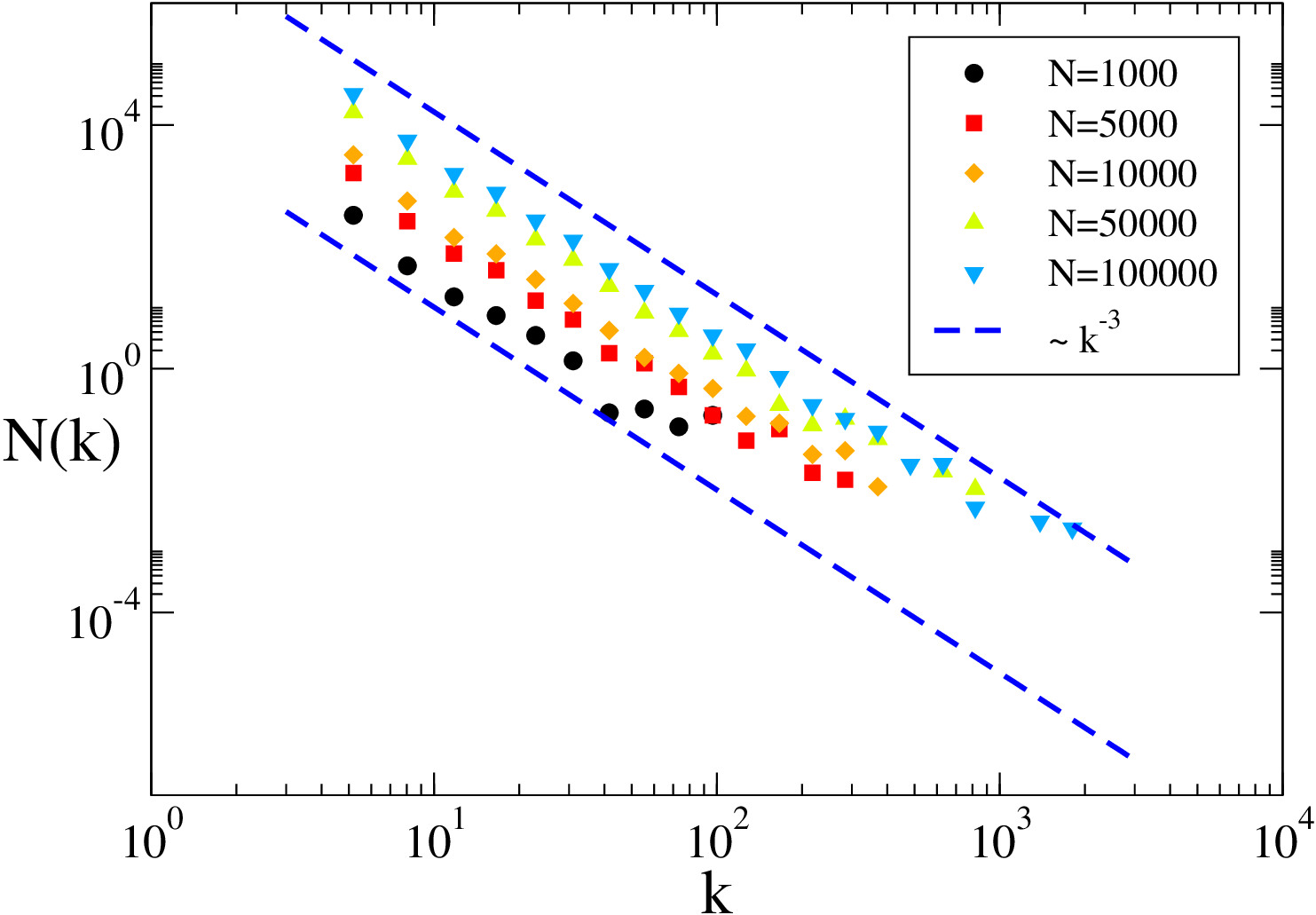}
  \end{center}
  \caption{\textbf{Degree distribution on the first layer for
      different values of $N$}. Even for relatively small multiplexes,
    i.e. $N=1000$, the exponent of the degree distribution is almost
    equal to $\gamma=3$.}
  \label{fig:suppl_fig_2}
\end{figure}

\begin{figure}[!ht]
  \begin{center}
    \includegraphics[width=5in]{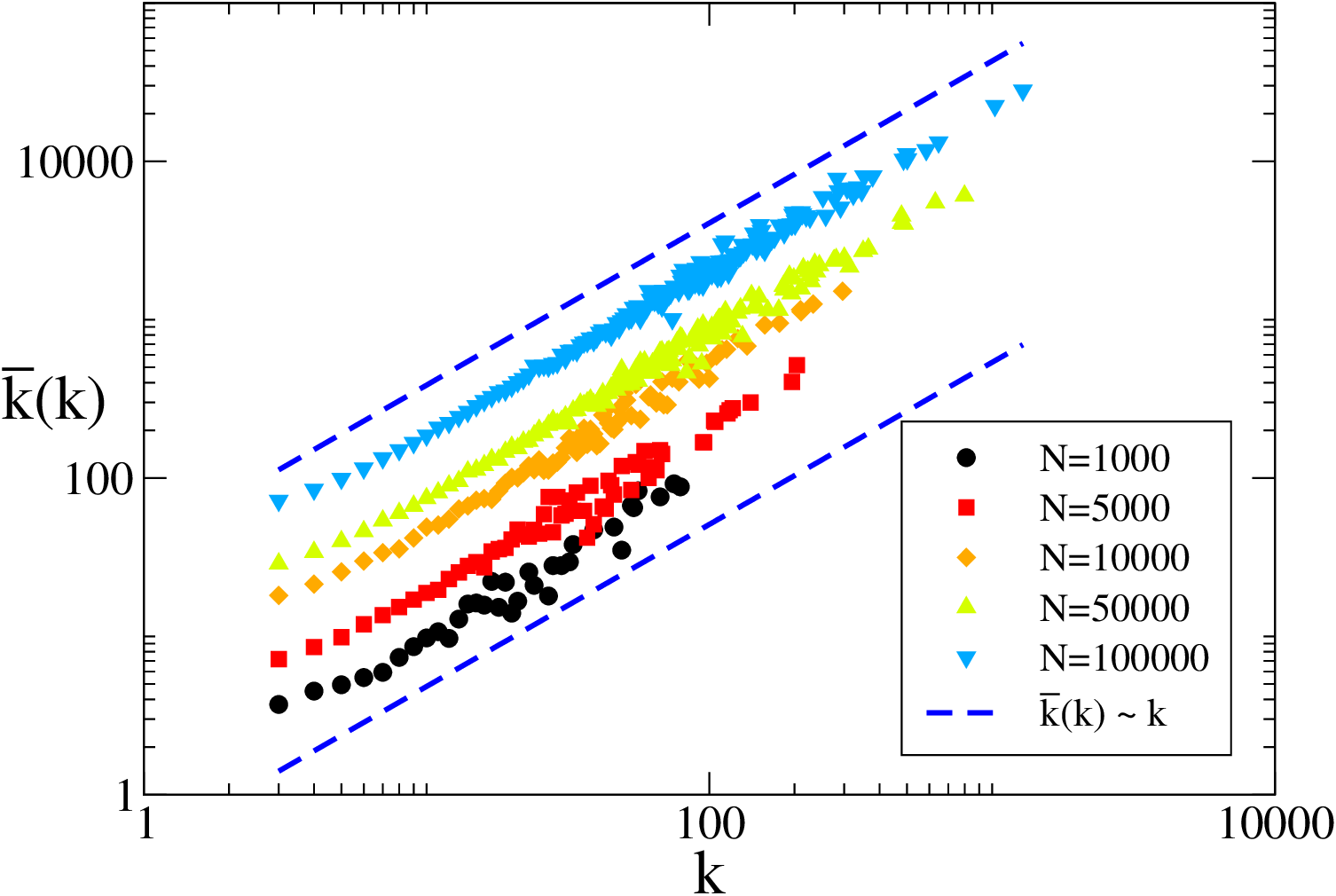}
  \end{center}
  \caption{\textbf{Inter-layer degree correlation as a function of
      $N$}. The shape of the inter-layer degree-degree correlations is
    already evident even for small values of $N$. The curves have been
    vertically displaced to facilitate visual comparison.}
  \label{fig:suppl_fig_3}
\end{figure}

It is interesting to investigate how the properties of the multiplexes
generated using the model we propose depend on the number of nodes
$N$. For instance, most of the mean-field predictions for the degree
distributions and inter-layer degree correlations are valid in the
limit of large $N$. However, as shown in Fig.~\ref{fig:suppl_fig_2}
and in Fig.~\ref{fig:suppl_fig_3}, the properties of the degree
distributions and of inter-layer degree-degree correlations are
similar to those predicted for large $N$ even for relatively small
multiplexes, e.g. with $N=1000$.

\section{Time complexity}

The most efficient algorithm for the construction of a simplex
networks based on preferential attachment takes advantage of random
sampling with rejection and runs in $\mathcal{O}(Nm^2)$. However, in
the case of a multiplex the procedure to sample a candidate neighbour
$j$ of a newly-arrived node $i$ is a bit more complicated. Let us
first consider a two-layer multiplex described by
Eq.~(\ref{eq:eq1}). When we sample the candidate neighbours of node
$i$ at layer $1$ at time $t$, each node $j$ should be sampled with a
probability proportional to $ak\lay{1}_j + (1-a)k\lay{2}_j$. The
simplest way to implement such sampling is to construct a vector
$\mathcal{S}$ whose $n$-th entry is equal to $\sum_{j=1}^{n}
ak\lay{1}_j + (1-a)k\lay{2}_j$ (the first element $\mathcal{S}[0]$ of
the array is set equal to zero); then we sample a real number $\zeta$
in the interval $\left(0,\mathcal{S}[t]\right]$ and we choose the node
  $j$ such that $\zeta\le \mathcal{S}[j]$ and
  $\zeta>\mathcal{S}[j-1]$. The construction of the vector
  $\mathcal{S}$ at each time $t$ requires $\mathcal{O}(t)$ operations
  while the sampling of a single node $j$ can be efficiently
  implemented by binary search, requiring at most
  $\mathcal{O}(\log(t))$ operations per edge, so that the sampling of
  $m$ edges requires at most $\mathcal{O}(m\log(t))$ steps. Thus, the
  total number of operations needed to sample a layer of a multiplex
  is:
\begin{equation}
  \sum_{j=m_0+1}^{N}\mathcal{O}(t) + \mathcal{O}(m\log(t)) =
  \mathcal{O}(N^2) + \mathcal{O}(mN\log(N)) =
  \mathcal{O}(N^2+mN\log(N))
\end{equation}
It is easy to verify that the construction of a $M$-layer multiplex
requires a number of steps
\begin{equation}
  \mathcal{O}(M(N^2+mN\log(N))
\end{equation}
which, for $m$ fixed, is dominated by $\mathcal{O}(M N^2)$. Therefore,
the time complexity of this algorithm is linear in the number of
layers and quadratic in the number of nodes. We notice that in
principle it is possible to construct better algorithms to sample
growing multiplexes by implementing a smart policy to update the array
$\mathcal{S}$.

\section{Randomly-chosen master layer}

\begin{figure}[!ht]
  \begin{center}
    \includegraphics[width=5in]{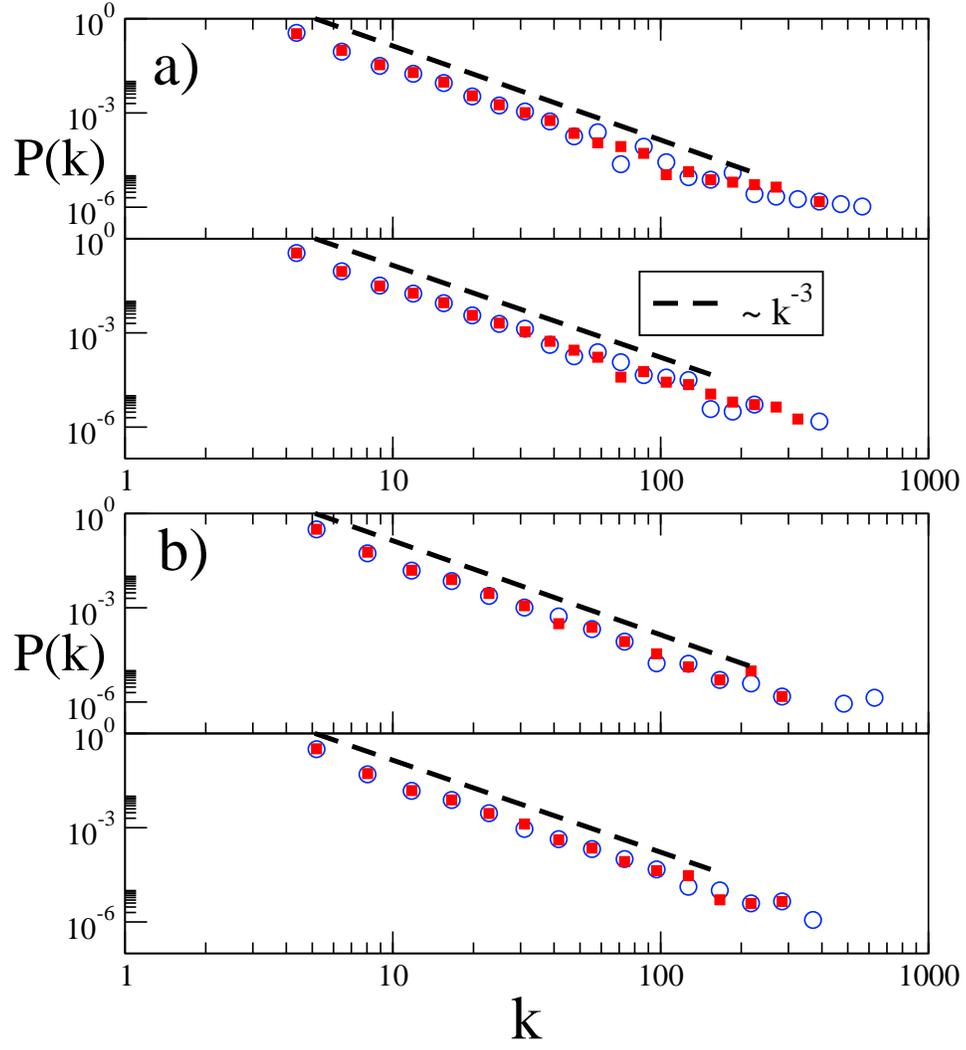}
  \end{center}
  \caption{\textbf{Degree distribution of mixing master layers}.  The
    degree distribution of the first layer for the EML (panel a) and
    the RML (panel b) for $\beta=1.1$ (blue circles) and $\beta=2.0$
    (red squares). The random choice of the master layer for each node
    does not sensibly change the shape of the degree distribution.}
  \label{fig:suppl_fig_4}
\end{figure}

\begin{figure}[!ht]
  \begin{center}
    \includegraphics[width=5in]{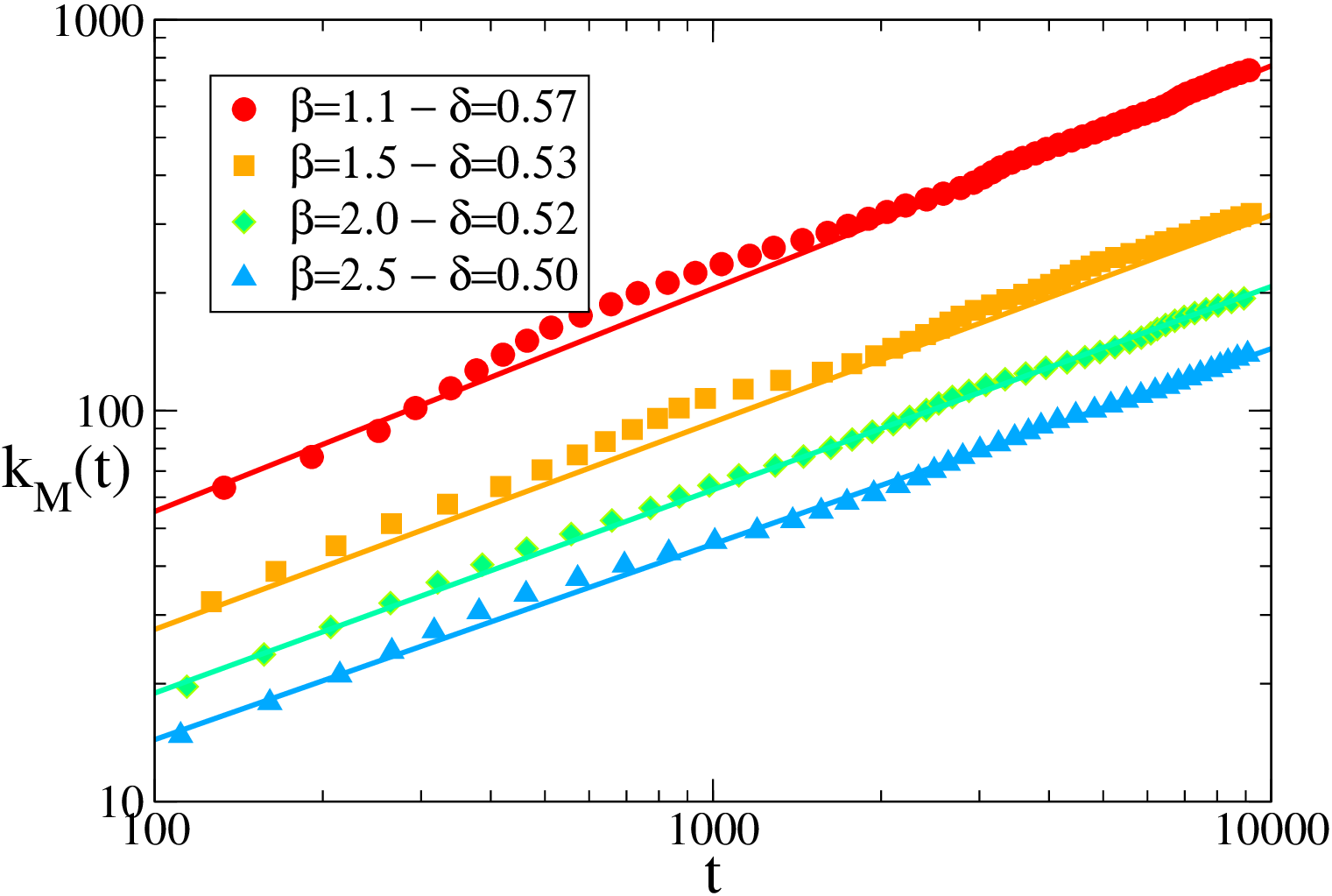}
  \end{center}
  \caption{\textbf{Temporal scaling of degree for RML}.  The degree
    $k_M$ of the largest hub in the first layer scales as
    $(t/s)^{\delta}$, where $\delta$ depends on the exponent $\beta$
    of the delay distribution and approaches $0.5$ as $\beta$
    increases. Notice that the values of $\delta$ are pretty similar
    to those observed in the case of EML, reportes in
    Fig.~\ref{fig:suppl_fig_1}.}
  \label{fig:suppl_fig_5}
\end{figure}

\begin{figure}[!ht]
  \begin{center}
    \includegraphics[width=5in]{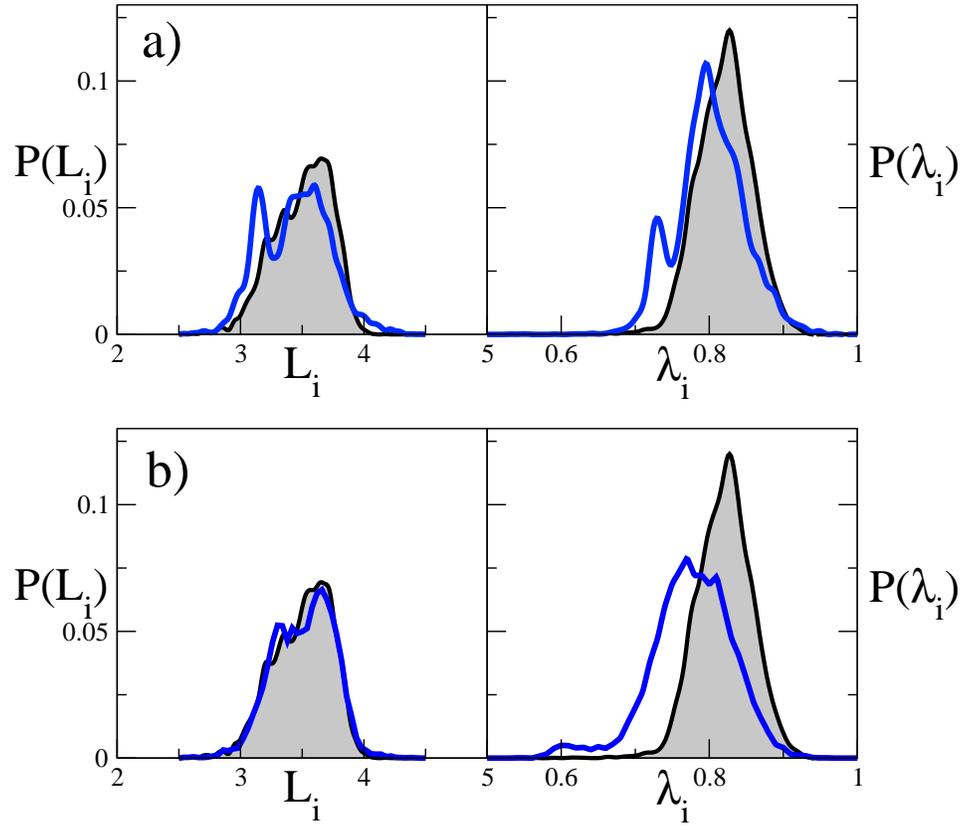}
  \end{center}
  \caption{\textbf{Distribution of shortest path lengths and node
      interdependence for RML}. When the master layer of each node is
    chosen at random, the distribution of shortest path lengths (panel
    b, left) is more similar to that obtained for synchronous arrival
    (shaded curve) than that obtained in EML (panel a,
    left). Conversely, the distribution of node interdependence for
    RML (panel b, right) sensibly deviates from that obtained for
    synchronous arrival (shaded curve) and differs from that observed
    in EML (panel a, right). These results can be explained by a more
    balanced distribution of super-hubs among the two layers of the
    multiplex.}
  \label{fig:suppl_fig_6}
\end{figure}

In the main text we made the simplifying assumption that each node
arrives first on the master layer and then on the other layers, after
a certain delay. We call this assumption ``Equal master layer''
(EML). In this Section we briefly comment on the case in which this
assumption does not hold, i.e. when a node first arrives either on the
first or on the second layer, and then arrives on the other layer
after a power-law distributed delay. We call this case
``Randomly-chosen master layer'' (RML), to stress the fact that the
master layer of each node is chosen at random among the $M$ layers of
the multiplex. In particular, we are interested in the case in which a
newly arrived node selects one of the $M$ layers of the multiples as
its master layer with uniform probability $p=1/M$. In
Fig.~\ref{fig:suppl_fig_4}, \ref{fig:suppl_fig_5} and
\ref{fig:suppl_fig_6} we report, respectively, the degree
distributions, the temporal scaling of the degree of the largest hub
and the distribution of shortest path lengths and node interdependence
for RML with $M=2$. The plots suggest that the random choice of the
master layer produces a more balanced distribution of super hubs
between the two layers, which has a relevant impact on the
distribution of shortest path lengths and node interdependence
(Fig.~\ref{fig:suppl_fig_6}). Conversely, the degree distributions and
the temporal scaling of the degree of the largest hub are practically
indistinguishable from those observed in EML
(Fig.~\ref{fig:suppl_fig_4} and \ref{fig:suppl_fig_5}).


\begin{thebibliography}{99}
\bibitem{footnote0} It is easy to show that the time required to
  simulate the growth of a multiplex with $N$ nodes and $M$ layers
  increases as ${\cal O}(N^2M)$. See Appendix for additional details
\bibitem{footnote1} Notice that for $t\gg1$ we have
  $\Pi_{k,q}^{[\alpha]}\ll1 \quad \forall k,q$; consequently,
  aymptotically in time we can neglect the probability that the same
  node acquires, at time $t$, a link in layer $1$ and in layer $2$.
\bibitem{footnote2} All the results shown in the figures correspond to
  2-layer multiplexes with $N=10000$ nodes, $m=3$, $m_0=3$, and are
  averaged over 50 realizations.
\end{thebibliography}
\end{document}